\documentclass[final,5p,times,twocolumn,authoryear]{elsarticle}

\usepackage{aas_macros}
\usepackage{amssymb}
\usepackage{lipsum}
\usepackage{amsmath}

\journal{New Astronomy}

\begin{document}

\begin{frontmatter}



\title{A temperature scale of 1$\sim$2\,eV in the mass-radius relationship of white dwarfs of type DA}


\author[first]{Jin Lim}
\author[first]{Ji-Yu Kim}
\author[first]{Maurice H.P.M. van Putten\corref{cor1}\fnref{label2}\fnref{label3}}
  \ead{mvp@sejong.ac.kr}
\fntext[label3]{INAF-OAS Bologna via P. Gobetti 101 I-40129 Bologna Italy, Italy}
\affiliation[first]{organization={Physics and Astronomy, Sejong University}, addressline={209 Neungdong-ro}, postcode={05006},
            city={Seoul},country={South Korea}
         }

\begin{abstract}
The mass-radius relationship of white dwarfs (WDs) is one of their defining characteristics, largely derived from electron degeneracy pressure.
We present a model-independent study of the observed mass-radius relationship in WD binaries of \cite{Parsons_2017}, listing data over a broad temperature range up to about 60,000\,K (5\,eV). The data show an appreciable temperature sensitivity with pronounced intrinsic scatter (beyond measurement uncertainty) for the canonical He-models with proton$-$to$-$neutron ratio 1:1. We characterize temperature sensitivity by a temperature scale $T_0$ in model-agnostic power-law relations with temperature normalized radius. For low-mass WDs, the results identify a remarkably modest $T_0 = 1 \sim 2 $\,eV. 
We comment on a potential interpretation for atmospheres insulating super-Eddington temperature cores from the sub-Eddington photospheres of low-mass WDs.
\end{abstract}




\end{frontmatter}




\section{Introduction}\label{introduction}

\begin{figure}[!t]
\vskip-0.3in
    \centerline{\includegraphics[scale=0.3]{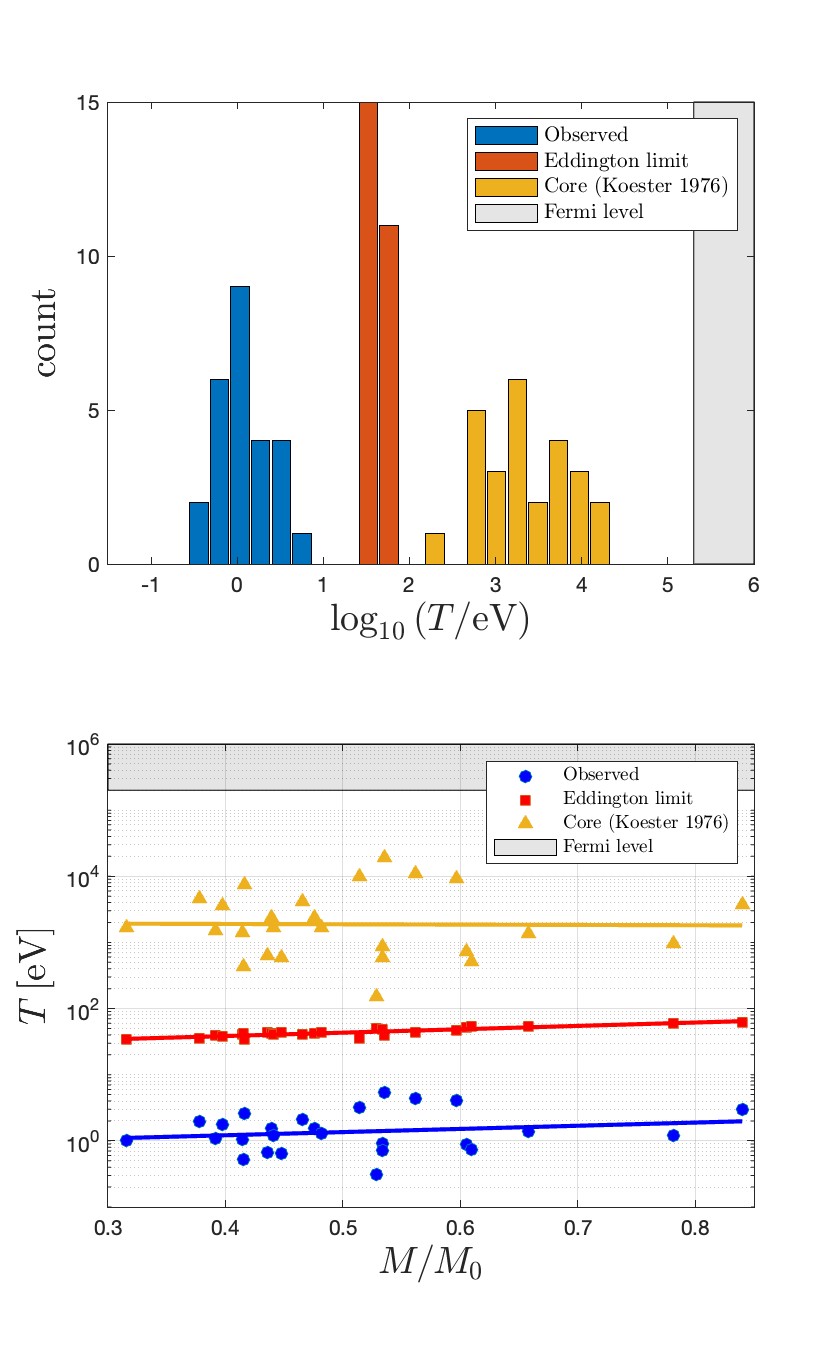}}
    \caption{(Top panel.) Observed, Eddington and core temperatures of the 26 WDs of type DA in the sample of \cite{Parsons_2017}. 
    Core temperatures are inferred from the \cite{koe76} correlation for an optically thick atmosphere, insulating a C/O or He core at super-Eddington temperatures. 
    (Lower panel.) Same data plotted as a function of mass with trends at slope 1.10 (Observed, {\em blue}), 1.18 (Eddington, {\em red}) and -0.11 (Core, {\em brown}). 
    Eddington temperatures are roughly consistent with
    the geometric mean of observed and core temperatures.
    }
    \label{figEDD}
\end{figure}

White dwarfs (WDs) represent the final evolutionary stage of stars with masses in the range of $0.4M_\odot<M<8M_\odot$. 
As such, they are quite numerous. The
Gaia Early Data Release 3 (EDR3) contains 359,073 WD candidates \citep{Fusillo21}. EDR3 extends the survey of 29,294 WDs of the {\em Sloan Digital Sky Survey} (SDSS) {\em Data Release 16} (DR 16) by over an order of magnitude with 25,176 WDs in both surveys. 
WDs are variously classified by their atmospheric composition (\citealt{Fusillo21}). 
Most are of type DA characterized by a dominance of hydrogen lines in their spectra. It reveals an opaque hydrogen atmosphere, whose observed temperatures are well below the Eddington limit (Fig. \ref{figEDD}).

WDs have a distinct mass-radius relationship with a lower bound defined by electron degeneracy pressure \citep{chandrasekhar1939book}. 
It is described by the equation of state (EoS) of degeneracy pressure of a Fermi gas, the relativistic limit of which defines the Chandrasekhar mass-limit.  
To leading order, WDs are modeled by the ideal Fermi gas at zero temperature surrounded by a non-degenerate atmosphere \citep{koester1990physics}.
At zero temperature, the degenerate WD core is effectively parameterized only by a proton-to-neutron ratio. 
In this limit, the mass-radius relation satisfies \citep{chandrasekhar1939book,hamada1961models,koester1979atmospheric}
\begin{eqnarray}
    R = C_0\left(\frac{M}{M_\odot}\right)^{-1/3},
\label{EQN_1}
\end{eqnarray}
where 
\begin{eqnarray}
    C_0 = \left( \frac{16G^3m_e^3}{81\pi^2\hbar^6}\right)^{-1/3} \left(\frac{Am_p}{Z}\right)^{-5/3} M_\odot^{-1/3}
    \simeq 0.01\,R_\odot ,
\label{EQN 2}
\end{eqnarray}
where $G$ is Newton's constant, $\hbar$ is Planck's constant, $m_e$ and $m_p$ are the masses of the electron and proton, $A$ is the atomic number and $Z$ denotes the number of protons in a nucleus - here, mostly alpha-nuclei. 
We have set the A/Z ratio to 2, since many WDs consist of a C/O or He core, the latter in particular expected for low-mass WDs \citep{iben1985evolution,marsh1995low,nelemans2001population,han2002origin,istrate2016models,ren2018white,zenati2019formation}. According to this calculation, $R = C_0M^{-1/3}$ holds. 

\begin{figure}[!t]
    \centerline{
    \includegraphics[scale=0.260]{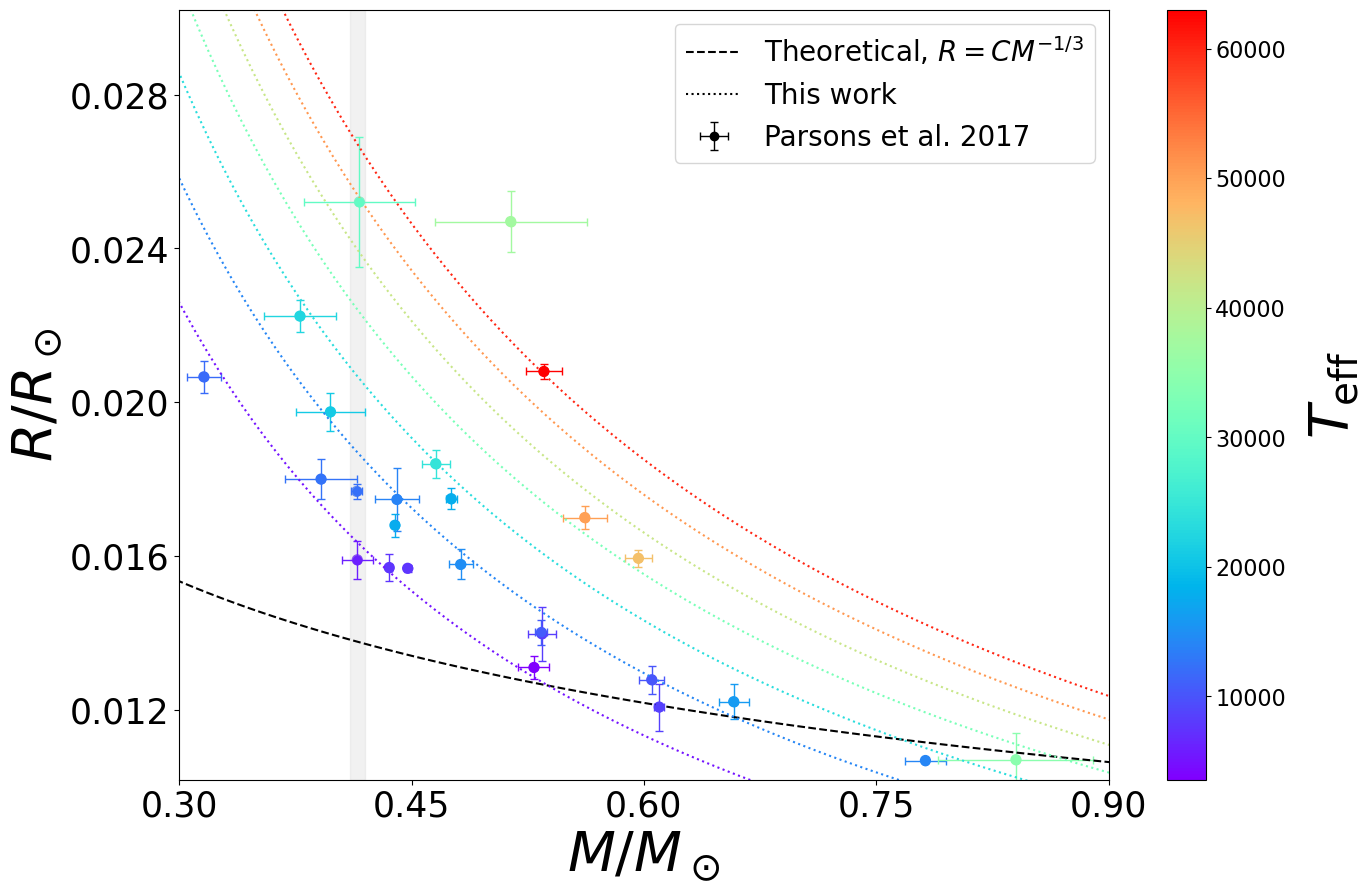}}  
    \caption{ 
    Mass-radius plot with temperature (color) in the sample of 26 WDs of type DA \citep{Parsons_2017}. 
    NN Ser is the hottest and SDSS J0138-0016 is the coldest at 63,000K and, respectively, 3,570 K. 
    For reference, it includes the theoretical zero-temperature limit (\ref{EQN_1}) (dashed black line). 
    The vertical grey region highlights three WDs of essentially the same mass with pronounced expansion in apparent radius with temperature. 
    A significant {\bf departure} is seen between observed and the expected radius (\ref{EQN_1}), 
    especially at low mass $M/M_\odot \lesssim 0.5$. Included are fits to the data by a model-agnostic power-law Relation-2 (dotted colored curves, \S3) studied in the present work with isothermals covering 5,000-65,000\,K (bottom to top in steps of 10,000\,K) in observed temperature. 
    {\bf Relation-2} identifies a characteristic temperature scale $T_0=1\sim 2\,$eV in the observed mass-radius relationship.}
    \label{fig.1}  
    \label{fig:color}
    \end{figure}

Fig. \ref{fig.1} shows the observed mass-radius relation for 26 WDs of type DA, all in eclipsing binaries, sampled by \cite{Parsons_2017} by photometric and spectroscopic observations. 
Photometric observations use, for example, the {\em Ultrafast Triple-beam CCD Camera} (ULTRACAM, \citealt{dhillon2007ultracam}) and its spectroscopic version ULTRASPEC \citep{dhillon2014ultraspec}, currently in use as the high-speed imaging photometer on the {\em Thai National Telescope} (TNT). Spectroscopic observations have been performed by X-shooter \citep{venet2011} in ESO {\em Very Large Telescope} (VLT).

Fig. \ref{fig.1} shows the theoretical mass-radius relation of the degenerate core (\ref{EQN_1}) to provide {a lower bound. It effectively provides a greatest lower bound only for $M/M_\odot \gtrsim 0.5$,} even though it represents an rather elementary model of degeneracy pressure. 
The data clearly show a temperature sensitivity {at relatively low mass $M/M_\odot \lesssim 0.5$, see also Fig. 9 in \cite{Parsons_2017}.} 
Crucially, the temperatures involved (Fig. \ref{figEDD}) are far below the characteristic energy $0.1-1$\,Me~V of the Fermi energies of the electrons in the degenerate core. For the present temperature range and low-mass WDs, any finite temperature corrections to the EoS \citep{carava2014,Bosh2018,Bosh2021}, 
including relativistic corrections, will be accordingly small for any generalizations beyond (\ref{EQN_1}), notably in \cite{hamada1961models}, \cite{rot11}, \cite{dec14}, \cite{Bosh2018} and \cite{bai19}; see further \cite{koe90,koe02}.
Moreover, at sufficiently high density, the mass-radius relationship of the core becomes universal, independent of the details of the EoS.

Instead, the origin of temperature sensitivity in the mass-radius relationship may be found in a non-degenerate atmosphere, if present. In particular, a finite temperature sensitivity is expected from an atmosphere about a core at super-Eddington temperatures - allowed for a sufficiently massive and optically thick atmospheres 
\citep{fontaine2001potential}.

A variety of studies have been conducted to find solutions thereto and gain a deeper understanding of, e.g., an H envelope and/or evolution models depending on core composition \citep{hamada1961models,hearn1976corona,verbunt1988mass,benvenuto1999grids,fontaine2001potential,panei2007full,boshkayev2015general,Parsons_2017,kepler2019white,pei2022highly}. Among these studies, various convection theories have been advanced, also to explain cooling times and evolutionary processes of WDs including the thermal insulation provided by their non-degenerate atmospheres. 

For instance, hot WDs in SDSS DR12 have been analyzed with models of cooling and atmospheres \citep{bedard2020spectral}. Non-Local Thermal Equilibrium (non-LTE) atmospheres and synthetic WD spectra reveal a correlation between surface gravity and effective temperature.

These modeled approaches, however, are intricate with potentially systematic uncertainties in the detailed structure of non-degenerate atmospheres, mediating heat transport by radiation and convection \citep{bed24}. For this reason, we set out the present model-agnostic study based on spectroscopic and photometric data, to further our understanding of the mass-radius relationship \citep{tremblay2016gaia,boshkayev2016equilibrium}.

For the recent sample of 26 WDs of \cite{Parsons_2017} (Fig. \ref{fig.1}), we set out to derive a temperature scale $T_0$, characterizing temperature sensitivity by exploring various power law relations for the mass-radius relations. The resulting $T_0$ may serve as a novel observational constraint in future studies.

In  \S2, we recall some preliminaries of the Eddington temperature and the \cite{koe76} correlation of core and observed temperature.
In \S3, we introduce a temperature normalized radius, to be used in power-law fits to the data based on two cost functions: $\chi^2$ and  residual Standard deviation (STD) defined by minimal least square errors. 
In \S4, introduce three temperature-normalized power-laws and consider their fits to data in the log-log plane to effectively describe the expansion of apparent radius with normalized temperature. 
In \S5, two of the three relationships are ranked by probability of significance by Monte Carlo analysis. 
In \S6, we interpret the results and summarize our findings with an outlook for future studies in \S7.

\section{The intermediate Eddington temperature}

Fig. \ref{figEDD} shows, as expected, the observed temperatures $T$ to be strictly below the Eddington temperature, $T_{Edd}$. After all, the modest observed temperatures on the order of a few eV imply the existence of an atmosphere. A reverse inequality would imply rapid evaporation by radiation pressure acting on the optically thin outer-most layers of any atmosphere.

The Eddington temperature $T_{Edd}$ is defined by equating the Eddington luminosity $L_{Edd}=3.2\times 10^4 \left(M/M_\odot\right)L_\odot$ to the luminosity from a sphere of radius $R$. That is, $L_{Edd}=4\pi R^2 \sigma T^4_{Edd}$, where 
$\sigma=5.67\times 10^{-5}{\rm g\,s}^{-3}{\rm K}^{-4}$ is the Stefan-Boltzmann constant. This defines
\begin{eqnarray}
    T_{Edd} = 39.5\,\left(\frac{g}{5000 \, g_\odot}\right)^\frac{1}{4}\,{\rm eV} 
\label{EQN_P1} 
\end{eqnarray}
by surface gravity $g=GM/R^2$ with a fiducial
value for $M=0.5M_\odot$, $R=2\% R_\odot$, 
scaled to the solar value $g_\odot = GM_\odot/R_\odot^2$. 
For the sample of \cite{Parsons_2017}, $T_{Edd}$ in (\ref{EQN_P1}) is on-average about 40 times the observed surface temperature $T$ (Fig. \ref{figEDD}). 

The Eddington temperatures shown are roughly consistent with the geometric mean $\sqrt{T_cT}$ of observed and core temperatures. As such, $T_{Edd}$ provides a natural reference to their correlations. 

Following a detailed revisit of WD envelopes for the observed temperature $T$ at the surface and the central temperature $T_c$ of the core of the WD, \cite{koe76} derives a correlation $T^4=2.05\times 10^{-10}\left(g/{\rm cm\,s}^{-2}\right)T_c^\alpha$ with index $\alpha=2.56$, where $T$ is in K. 
Scaled to $T_{Edd}$, it takes the form
\begin{eqnarray}
T_c \equiv  \eta T_{Edd}
\label{EQN_P2}
\end{eqnarray}
with
\begin{eqnarray}
\eta = 35.7\, R_{0.02}^{1/2}\,M_{0.5}^{-1/4} \left(\frac{40\, T}{T_{Edd}}\right)^\beta,
\label{EQN_P3}
\end{eqnarray}
$\beta=4/\alpha\simeq 1.56$. Here, we use the notation $R=2\% \,R_{0.02}\,R_\odot$ and $M=0.5\,M_{0.5}M_\odot$ for a fiducial value and taking into account aforementioned mean ratio of 
$T_{Edd}$ to $T$. 

Fig. \ref{figEDD} summarizes the distributions of $T_c$, $T_{Edd}$ and $T$ for the sample of \cite{Parsons_2017}.
These are well below the Fermi level $E_F$ of the degenerate electrons supporting the core with characteristic temperature $k_BT_c=z \,m_pc^2\sim0.1-1$\,MeV, where $z=R_g/R$ is the gravitational redshift of the WD surface according to its gravitational radius $R_g=GM/c^2$, where $m_p$ is the proton mass, $k_B$ is the Boltzmann constant, and $c$ is the velocity of light. 

\section{Temperature normalized radius}

Fig. \ref{fig.1} shows the data in a mass-radius plot alongside the theoretical zero-temperature limit (\ref{EQN_1}). Highlighted by color is a general trend of increasing radius with temperature. This trend is particularly striking upon considering similar masses. 
Table 1 lists three WDs of mass $M \simeq 0.415 M_\odot$ clearly showing a pronounced correlation of apparent radius expanding by $\sim50\%$ with temperature increasing to $\sim 5\,$eV. 

In absolute terms, relative to the Fermi level of the electrons (Fig. \ref{figEDD}), these temperatures are extremely modest leaving the star essentially unperturbed \citep{hamada1961models}. 
{For the present sample of \cite{Parsons_2017}, this suggests, instead, a temperature sensitivity in the H atmosphere of WDs considered previously by modeled approaches in \cite{Parsons_2017} more likely so than in the degenerate core. We return to this in \S7.}

Here, we circumvent model assumptions by using generic and model-agnostic power-laws for an effective description by a temperature-normalized radius. 
In doing so, we derive a characteristic temperature scale characterizing temperature sensitivity, blind to the underlying physical origin.
To be specific, based on Fig. 1 and Table 1, we consider 
\begin{eqnarray}
    R=R_0\left(\frac{M}{M_\odot}\right)^\alpha f(T/T_0)
    \label{EQN 3}
\end{eqnarray}
with free parameters $\left(\alpha,T_0\right)$. Here, $f(T/T_0)$ is dimensionless and $R_0$ is a constant fixing the dimension of length.

Starting point of our approach are effective mass-radius relationships of the form $R \propto M^\alpha f(T/T_0)$ for some power-law index $\alpha$ and temperature scale $T_0$. Equivalently, this considers a correlation of mass to the scaled radius
\begin{eqnarray}
R^\prime  = \frac{R}{f(T/T_0)},
\label{EQN 4}
\end{eqnarray}
We apply the temperature scaled radius (\ref{EQN 4}) to 
fit the data in the form
\begin{eqnarray}
    R^\prime = R_0 \left(M/M_\odot\right)^\alpha.
\label{EQN 5}
\end{eqnarray}
For the sample of \cite{Parsons_2017}, we determine best-fit parameters $\left(\alpha,T_0\right)$ to the mass-radius data of 26 WDs of type DA, all in eclipsing binaries. 
\begin{table}
\centering
\caption{Three WDs of essential the same mass showing a clear increase of observed radius with temperature in the sample of \cite{Parsons_2017}.}
\vspace{0.2cm}
\renewcommand{\arraystretch}{1.5} 
\resizebox{0.49\textwidth}{!}{  
    \begin{tabular}{|c|c|c|c|c|} \hline 
         Object & $M\,[M_{\odot}]$ &  $R\,[R_{\odot}]$ & $T_{\rm eff}\,$[K] & $k{_B}T_{\rm eff}\,$[eV] \\ \hline 
         CSS 0970 & 0.4146 &  0.025 & 30000 & 2.9 \\ \hline 
         SDSS J1028+0931 & 0.4146 &  0.018 & 12000 & 1.1 \\ \hline 
         SDSS J1210+3347 & 0.4150 &  0.016 & 6000 & 0.52 \\ \hline
    \end{tabular}
}
\label{table.1}
\end{table}

The function $f(T/T_0)$ in (\ref{EQN 3}), to be discussed further below, will be a power-law comprising the free parameters $\left(\beta,T_0\right)$. 
In a fit to the mass-radius data, these parameters will be considered over a broad range of values
\begin{eqnarray}
     0 < T_0 < 9 ~ \mbox{eV}, ~~0 < \beta < 1.
    \label{EQN 6}
\end{eqnarray}

The characteristic temperature scale $T_0$ is limited to $T_0<9\,$eV. 
In this energy range, radii obtain accurately, while beyond, accuracy diminishes. We keep $\beta<1$, reflecting the assumption that temperature has a secondary impact on the radius. Our best-fit is defined by $\alpha$ estimated using ODR (Orthogonal Distance Regression) and, subsequently, the optimal value of $\beta$ and $T_0$ at the minimum STD and $\chi^2$ of residuals.

{Following standard practice, our power-laws (\ref{EQN 3}-\ref{EQN 5}) are analyzed by fits to linear trends in the log-log plane to}
\begin{eqnarray}
    \log R^\prime  = \alpha \log M + C.
    \label{EQN_7}
\end{eqnarray}
In fits to data by (\ref{EQN_7}), \,$\chi^2$ optimizes both the index $\alpha$ in scaling by a power-law in mass and the constant $C$,
while minimizing STD optimizes only $\alpha$. 
{In the present analysis quantifying the goodness-of-fit according to residual scatter about a trend line (\ref{EQN_7}), the level shift $C$ along the ordinate - absorbing $R_0$ in (\ref{EQN 3}-\ref{EQN 5}) - is safely ignored and it suffices to determine $\alpha$ by minimization of residuals. } 

\section{Temperature-normalized mass-radius relations}

In the present model-agnostic approach, we explore fits to the data using three temperature-normalized power-laws.
To this end, optimize by
$\chi^2$ and STD in our parameter estimation from fits
to the sample of \cite{Parsons_2017}.

The first power-law {\em Relation-1} is
\begin{eqnarray}
    {\rm I.}\, f(x) = x^\beta,
    \label{EQN 8}
\end{eqnarray}
where $x=T/T_0$. Here, $T_0$ acts as a constant because of (\ref{EQN 6}). For this reason, $\beta$ is not affected by a choice of $T_0$. 
We find
\begin{eqnarray}
\begin{array}{lll}
\chi^2:     & \alpha = -0.954, \,\beta=0.195\\
{\rm STD}:  & \alpha = -0.951, \,\beta=0.190
\end{array}
\label{EQN_9}
\end{eqnarray}
with the minimum $\chi^2 = 0.0183$ and, respectively,
with residual $\sigma =  0.03664$.
The second power-law {\em Relation-2} is
\begin{eqnarray}
    {\rm II.}\,f(x) = \left(1+x\right)^\beta.
    \label{EQN_11}
\end{eqnarray}
In contrast to Relation-1, Relation-2 includes the zero temperature limit of the Chandrasekhar mass-radius relation (\ref{EQN_1}). 
Accordingly, $T_0$ is no longer ignorable and is determined in the optimization process in fitting (\ref{EQN_11}) to the data. We find
\begin{eqnarray}
\begin{array}{llll}
\chi^2:    & \alpha = -0.969, \,\beta = 0.389, \,T_0 = 16226\\
{\rm STD}: & \alpha = -0.965, \,\beta = 0.356, \,T_0 = 13896
\end{array}
\label{EQN_12}
\end{eqnarray}
with the minimum $\chi^2 = 0.0159$ and, respectively, 
with residual $\sigma =  0.0344$.
The third power-law {\em Relation-3} is
\begin{eqnarray}
    {\rm III.}\,f(x) = 1+x^\beta,
    \label{EQN 14}
\end{eqnarray}
similar but not identical to Relation-2. 
Relation-3 also includes the Chandrasekhar limit of zero temperature and $T_0$ is not ignorable. We find
\begin{eqnarray}
\begin{array}{llll}
\chi^2:    & \alpha = -0.971,\, \beta = 0.690, \,T_0 = 66500,\\
{\rm STD}: & \alpha = -0.965,\, \beta = 0.647, \,T_0 = 65914
\end{array}
\label{EQN_15}
\end{eqnarray}
with the minimum $\chi^2 = 0.0161$ and, respectively, 
a residual $\sigma =  0.0346$.

\begin{figure*}[!t]
    \centerline{
    \includegraphics[scale=0.25]{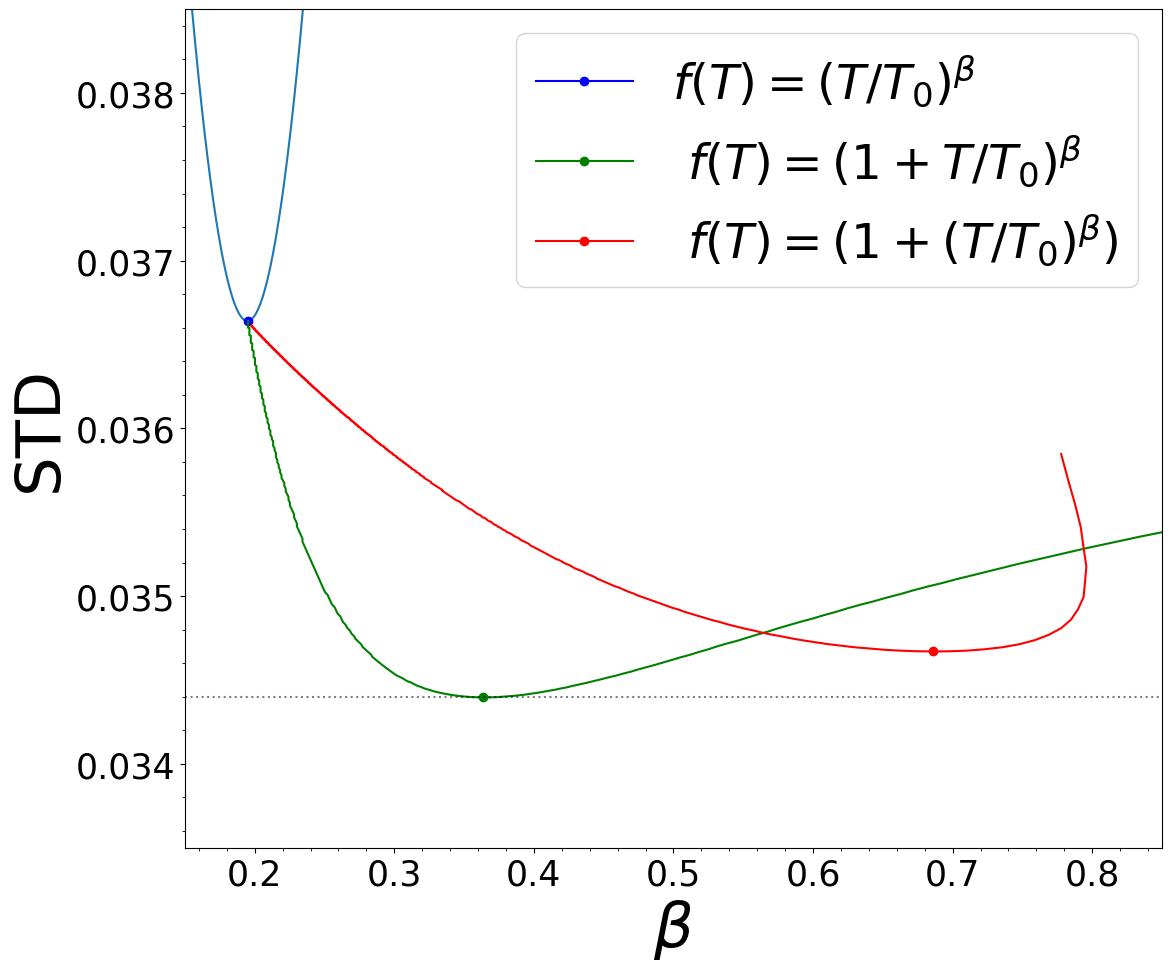}
    \includegraphics[scale=0.25]{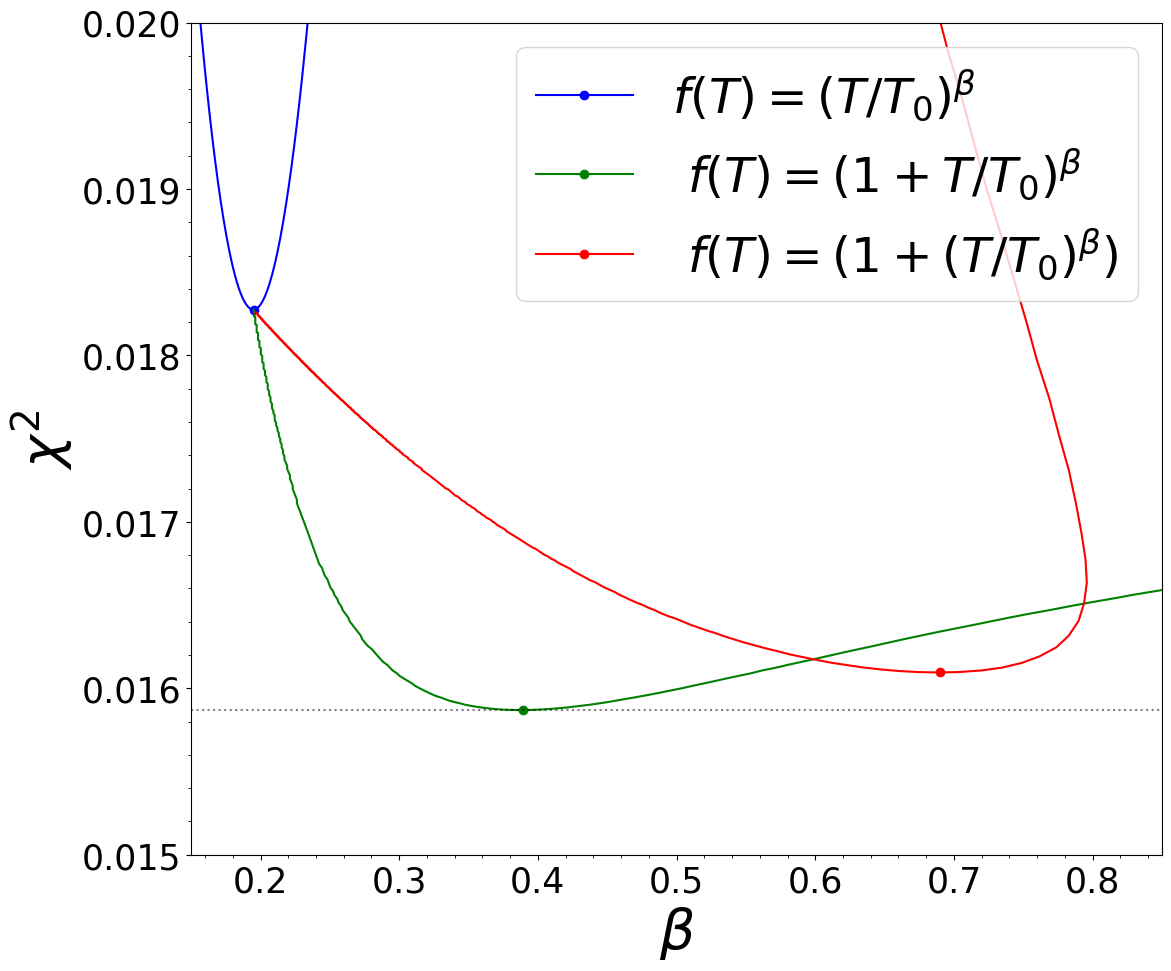}}
    \caption{Tracks of STD residuals (left panel) and $\chi^2$ (right panel) 
    for the Relation-1 (blue), Relation-2 (green) and Relation-3 (red) as a function of the normalized temperature power-law index $\beta$, following minimization over all $T_0$ in (\ref{EQN 6}). Note that Relation-1 has no explicit dependence on $T_0$. Both cost functions produce very similar results for $\alpha$ and $\beta$.}
    \label{fig.2}
\end{figure*}

Fig. \ref{fig.2} summarizes our three results, each indicated by color: blue, green and red for Relation-1, Relation-2 and Relation-3, respectively. 
The curve represents the optimal $\beta$ with each $T_0$ at the minimum STD ($\chi^2$).
The junction point shows the common minimum STD ($\chi^2$). 
Both Relation-2 and Relation-3 leave rather similar residuals (in $\sigma$ and $\chi^2$) for each of Relation-2 and Relation-3.
To rank these two relations, we proceed as follows.

\section{Ranking relations by Monte Carlo Analysis}

In this section, Relation-2 and Relation-3 are ranked for significance by Monte Carlo (MC) analysis.

MC analysis is a useful method for robust parameter estimation and ranking relations in the face of measurement uncertainties. 
Though ODR methods can be used also to estimate parameters, the results do not necessarily agree, making it difficult to rank Relation-2 and Relation-3 for their relative significance.

\begin{figure}[t]
    \centerline{
    \includegraphics[scale=0.35]{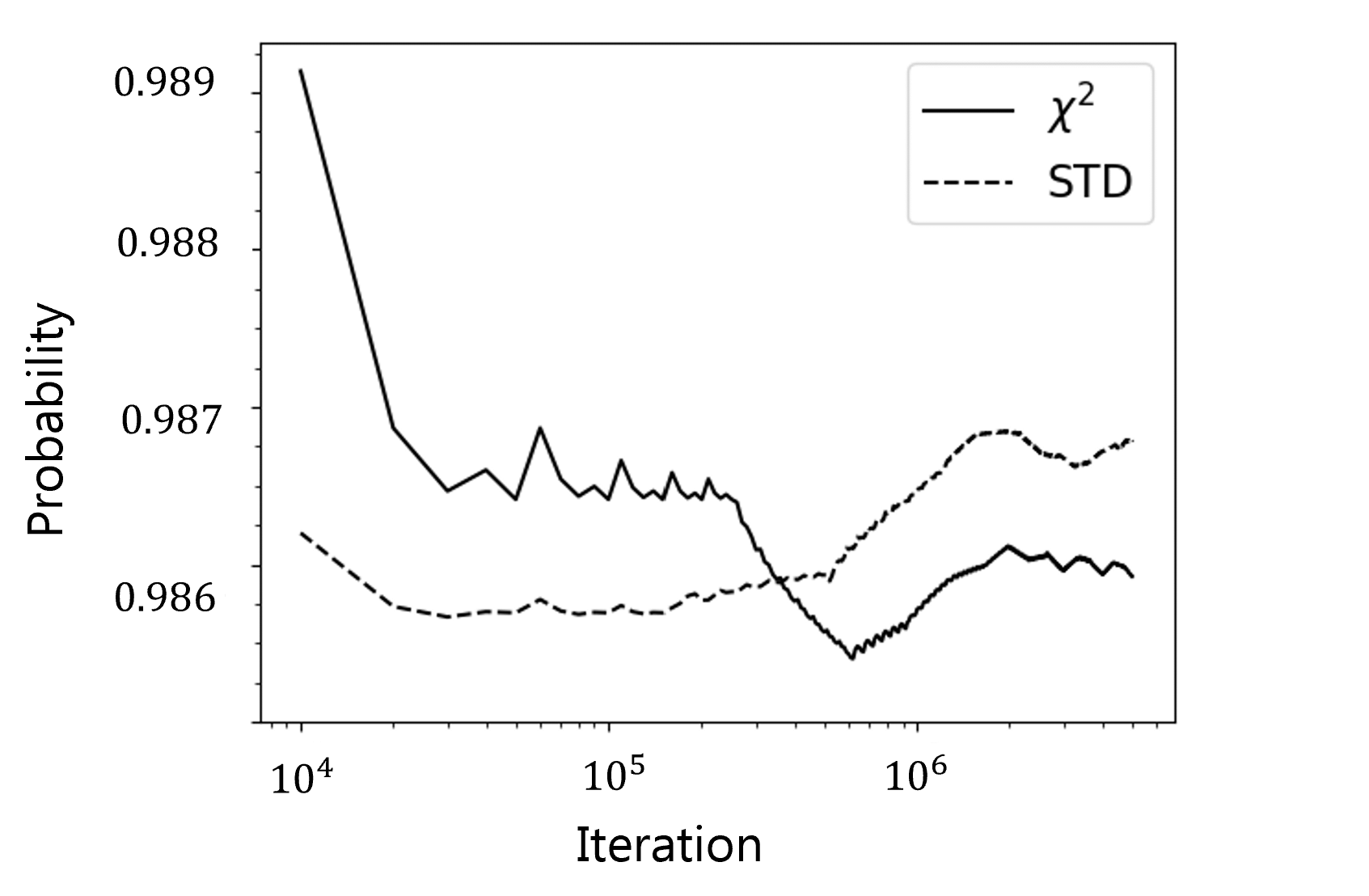}}    \caption{Probability of Relation-2 to have a smaller residual in $\chi^2$ (continuous curve) and STD (dashed curve) than Relation-3 in a MC analysis extending over a large number of synthetic data sets. Both cost functions support the conclusion that Relation-2 is preferred over Relation-3 in providing a model-agnostic fit to the mass-radius data of WDs.}
    \label{fig:4}
\end{figure}

In this light, we pursue MC analysis by creating synthetic data by randomly selecting samples of varying radius, mass, and temperature within the measurement confidence intervals,
\begin{equation}
    X_\text{{new}} = X + \mathcal{N}(0,\sigma).
    \label{eq.new_Data}
\end{equation}
Following this procedure, we estimate $\alpha$, $\beta$, and $T_0$ without considering errors in ODR. 
Results extended over 5M calculations are used to determine a ranking of Relation-2 and Relation-3 according to STD or $\chi^2)$ (Fig. \ref{fig:4}) and infer a probability $P$ of relative significance by counting the total number of times either one has preferred rank (Table \ref{tab_2}). The MC simulation also provides accurate values in the mean of $\alpha$, $\beta$, and $T_0$ over the total number of iterations.

In our MC analysis, {we consider realizations of arrays of $26\times 3 = 78$ entries}, comprising mass, radius and temperature of the 26 WDs. As observed quantities, mass, radius and temperature data are independent. An accordingly fair (unbiased) draw of realizations extends over random draws from 98 confidence intervals, unconstrained and independently, blind to physical meaning and pre-conceived notions of correlations. In our analysis, the range of allowed values is densely covering by using a very large number of 5M iterations.

Table \ref{tab_2} shows the output of our MC analysis.
The results indicate that the WD radius is firstly determined by mass more so than temperature. Relation-2 and Relation-3 have similar $\alpha$ values, but $\beta$ and $T_0$ are notably distinct with otherwise similar residuals in $\chi^2$ and STD (Figs. \ref{fig:color}-\ref{fig:3}).{Table \ref{tab_2} includes Relation-1, results for which are consistent with the above discussion.}

\begin{table}
\centering
\caption{Monte Carlo analysis on Relations 1-3 with ranking
by probability $P$ of having the lowest residual in $\chi^2$ (left) or STD (right) for the same data (Fig. \ref{fig:4}, synthesized over 5M representations).}
\vspace{0.2cm}
\renewcommand{\arraystretch}{1.5} 
\resizebox{0.5\textwidth}{!}{  
    \begin{tabular}{|c|c|c|c|c|c|c|} \hline 
    &  \multicolumn{3}{|c|}{$\chi^2$}&  \multicolumn{3}{|c|}{STD}\\ \hline 
    &  Relation-1&Relation-2 & Relation-3 &  Relation-1&Relation-2 & Relation-3 \\ \hline 
    $R_0/R_\odot$ &  $7.6 \times 10^{-3}$&$6.2 \times 10^{-3}$ & $5.8 \times 10^{-3}$ &  $8.4\times 10^{-4}$&$6.2 \times 10^{-3}$ & $5.8 \times 10^{-3}$ \\ \hline  
    $\alpha$ &  -0.951&$-0.966$ & $-0.967$ &  -0.951&$-0.965$ & $-0.966$ \\ \hline  
    $\beta$ &  0.194&$0.409$ & $0.690$ &  0.193&$0.375$ & $0.685$ \\ \hline  
    $\bar{T}_0$ &  25000&$19000$& $65000$&  const&$15000$& $64000$\\ \hline  
    $P$ & $\ll 1\%$ & $98.7\%$ & $\lesssim 1\%$ & $\ll 1\%$ & $98.6\%$ & $\lesssim 1\%$ \\ \hline 
    \end{tabular}
}
\label{tab_2}
\end{table}

\section{Interpretation of Results}


\begin{figure*}[!t]
    \centerline{\includegraphics[scale=0.25]{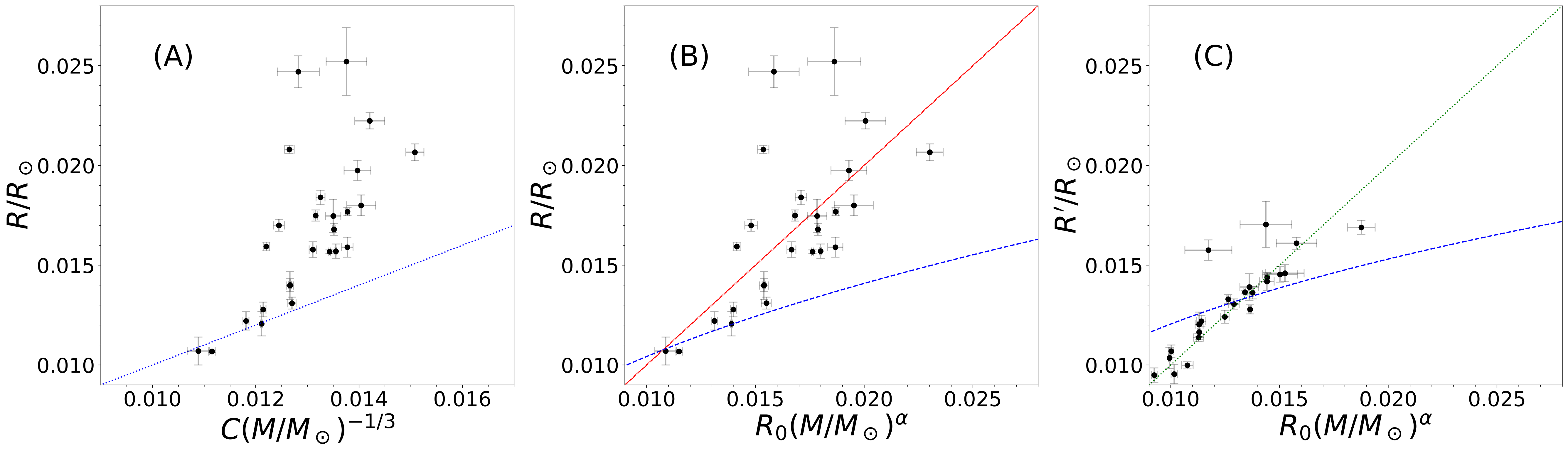}}
    \vskip0.2in
    \centerline{\includegraphics[scale=0.39]{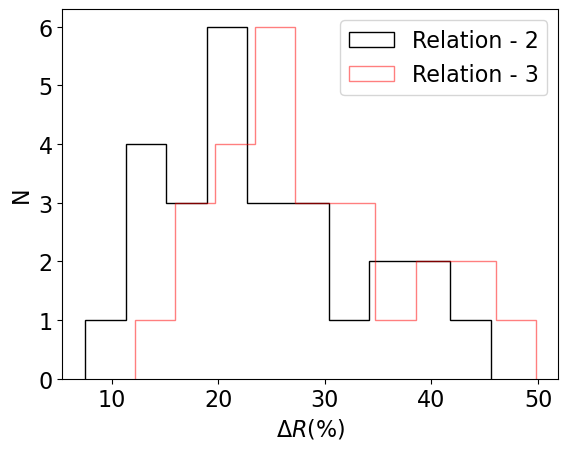}
    \includegraphics[scale=0.39]{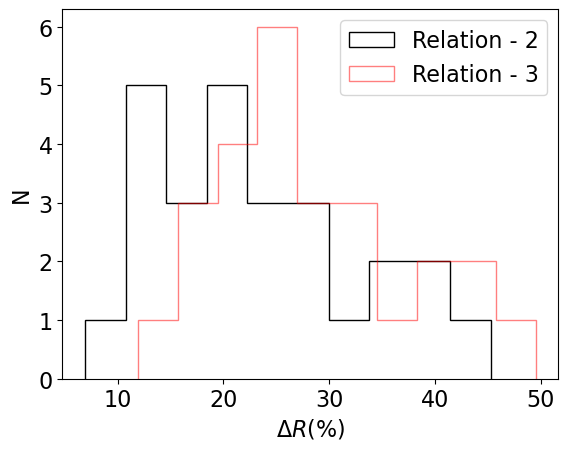}}
    \caption[A mass-radius relationship \label{fig:3}]{
    Mass-radius plots of the data (black dots). 
    Panel (A) highlights the deviation from the theoretical zero-temperature limit (\ref{EQN_1}). Panel (B) highlights the excess scatter in the data following a fit (red line) to the un-normalized radius $R \sim M^{\alpha}$. Panel (C) shows a fit (green line) to our temperature-normalized radius $R^\prime \sim M^\alpha$ in Relation-2 (\ref{EQN_11}). 
    A small residual scatter (cf. Fig. \ref{fig.2}) evidences the effectiveness of our normalization in Relation-2, except for three outliers with relatively large observational uncertainties.
    All three panels A-C include the theoretical relation (blue dotted line).
    Bottom panels show the adjusted radius in our normalization produced by minimization of $\chi^2$ (left) and STD (right). 
    The adjustment by our temperature normalization in $R^\prime = R/f(x)$, $x=T/T_0$, is about 23\% ($\sigma = 1.76$) 
    and 28\% ($\sigma=2.05$) in Relation-2 and, respectively, Relation-3 in $\chi^2$ optimization. The same is 24\% ($\sigma=1.82$) and 29\% ($\sigma= 2.07$) in Relation-2, respectively, Relation-3 in STD optimization. This adjustment is most relevant at high temperatures.}
    \label{fig:3}
\end{figure*}

Eclipsing binaries allow precise measurements of WD mass and radius \citep{Bours_2016,Parsons_2017}. 
{However, as in Fig. \ref{fig.1}, the theoretical mass-radius relation (\ref{EQN_1}) provides a lower bound. It is generally not the greatest lower bound by significant departures for hot, low-mass white dwarfs ($M/M_\odot \lesssim 0.5$).}
The gray region in Fig. \ref{fig.1} (Table \ref{table.1}) {is illustrative, highlighting a pronounced trend in observed WD radius with temperatures at otherwise very similar masses in $M/M_\odot<0.5$.} 
Evidently, this trend cannot be explained by the zero-temperature mass-radius relationship. 

Several theoretical calculations \citep{carava2014, Bosh2018, Bosh2019} have been advanced to explain this departure. While rotation and density affects the radius, the radius of relatively dense WDs is not significantly influenced by temperature even though such is more so for low-density or rotating WDs compared to their high-density counter parts.
Our a model-agnostic study of $T_0$ identify a characteristic temperature in the expansion of the  radius of the photosphere. It reveals a consistent trend wherein WDs with higher temperatures exhibit relatively larger radius, clearly apparent in the overall trend \citep{benvenuto1999grids,panei2007full,Parsons_2017,Joyce_2018,zenati2019formation,Romero_2019}.

In a novel model-agnostic revisit of the mass-radius relationship, we quantify this temperature dependence by a temperature scale $T_0=\left(1-2\right)$\,eV in Relation-2. 
Relation-2 is found to be statistically more significant than Relation-3 based on our MC analysis (\S4).
Over 5M iterations, Fig.\ref{fig:4} shows Relation-2 to have a lower $\chi^2$ and lower $\sigma$ than Relation-3 at a probability of 98.7\%, respectively, 98.6\% .
In the present approach, we circumvent potential systematic uncertainties otherwise present in model-dependent approaches (\citealt{Parsons_2017} and references therein).  

We summarize our approach in Fig. \ref{fig:3} (A-C). Panel (A) shows the discrepancy between the theoretical mass-radius relation at zero temperature (\ref{EQN_1}) and the observed radius. 
Panel (B) shows a fit to the mass to the un-normalized radius, 
revealing  scatter than clearly exceeds that of measurement uncertainty. 
Panel (C) shows the result of a linear relation between the temperature-normalized radius $R^\prime$ and $(M/M_\odot)^\alpha$. 
At relatively small residual scatter, this result identifies a temperature scale $T_0$ and a radius primarily determined by mass. 
According to Table 2 and in the notation of  ({\ref{EQN_1}}), we infer a mass-radius relation 
\begin{eqnarray}
    R = C\left(\frac{M}{M_\odot}\right)^{-1/3}
\label{EQN_1b}
\end{eqnarray}
with the temperature-dependent expansion of an atmosphere included in
\begin{eqnarray}
    C\simeq C_0\left(1+\frac{T}{T_0}\right)^{1/3}\,\left(\frac{M}{M_\odot}\right)^{-2/3}
\label{EQN_1c}
\end{eqnarray}
according to Relation-2, parameterized by STD in the approximations $\alpha=-0.965\simeq -1$ and $\beta=0.375\simeq 1/3$.

\section{Conclusions and Outlook}

A principal outcome of our model-independent study is a temperature scale $T_0=1\sim 2\,$eV in temperature sensitivity of the photospheric radius of {the WDs of type DA in \cite{Parsons_2017},} shown in Fig. \ref{fig:3} and summarized in (\ref{EQN_1b}-\ref{EQN_1c}). 

\begin{figure}
\centerline{\includegraphics[scale=0.25]{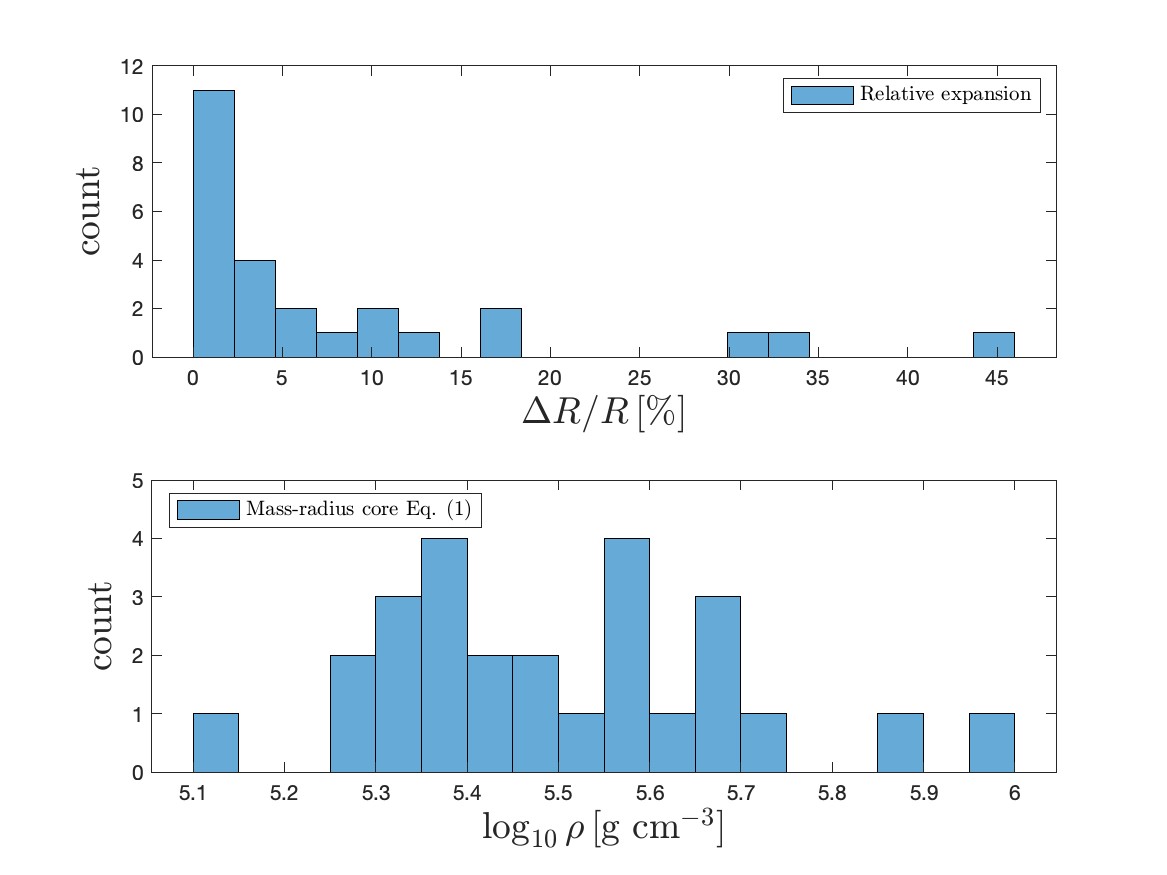}}
\caption{(Top panel.) For the WD sample of \cite{Parsons_2017}, shown are the expected relative expansion of a non-degenerate atmosphere,  sufficiently massive and opaque, surrounding a degenerate core based on super-Eddington temperature $T_c$ inferred from the \cite{koe76} correlation to the observed temperature $T$ (Fig. 1). 
The \cite{Parsons_2017} sample satisfies $T_{Edd}\sim\sqrt{T_cT}$, reconciling $T_c\gg T_{Edd}$ with $T\ll T_{Edd}$, leaving $T$ on the order of a few eV. (Lower panel.) For the entire \cite{Parsons_2017} sample, the mass-densities of the core according to the theoretical zero-temperature relationship (1) are above the threshold $10^5$\,g\,cm$^{-3}$ of \cite{Bosh2018}, where temperature sensitivity is negligible.}
\label{fig6}
\end{figure}

{While the \cite{Parsons_2017} sample of WDs covers a sizeable range in masses with distinct temperature sensitivity below $M/M_\odot \lesssim 0.5$ and above $M/M_\odot \gtrsim 0.5$, they nevertheless are of relatively high mass-density $\rho\gtrsim \rho_0$ (Fig. \ref{fig6}). Crucially, densities for the entire at hand are above the threshold $\rho_0=10^{5}$\,g\,cm$^{-3}$ above which the mass-radius relation of the degenerate core is expected to be insensitive to temperature \citep{Bosh2018}. 
In this sense, the present sample is relatively homogeneous and does not test temperature sensitivity of the core beyond what is expected from (\ref{EQN_1}).} 
  
{Our model-agnostic analysis hereby effectively reveals the presence of non-degenerate atmosphere, sufficiently massive and opaque, to account for the observed temperature sensitivity in Fig. \ref{fig.1} parameterized by above-mentioned $T_0$.
{Indeed, further confirmation can be found in consistency of the present \cite{Parsons_2017} data with the detailed model for H atmospheres of \cite{bedard2020spectral}.
} 

We summarize the apparent mass-radius relation (\ref{EQN_1b}) with temperature dependent coefficient (\ref{EQN_1c}) due to this atmosphere by including Relation-2 as a factor modifying $C_0$ in (\ref{EQN_1c}).}

$T_0$ derives from $1.6\,$eV and $1.3\,$eV according to $\chi^2$ and, respectively, STD in fits of Relation-2 to the data. $T_0$ appears to be particularly relevant to low mass $M/M_\odot\lesssim 0.5$ in the present sample.

For the above-mentioned low-mass WDs, a non-degenerate atmosphere effectively insulates a core at super-Eddington temperatures (\ref{EQN_P2}) from the observed sub-Eddington temperatures $T$ by virtue of an atmosphere that is sufficiently massive and opaque. 
In the following scaling, we shall ignore a potential role of a corona due to magnetic fields \citep{Aznar_Cuadrado_2004,Jordan_2006,Ferrario_2020}. 

Following \S2, the Eddington temperature introduces a scale height $h_E$ by virtue of thermal kinetic energy of the ions, distinct from escape by radiation pressure. For a hydrogen atmosphere, we have
\begin{eqnarray}
h_E\equiv \frac{k_BT_{Edd}}{m_pg}
=13\,{\rm km}\,R_{0.02}^2\,M_{0.5}^{-1}
\simeq 0.3\% R_{WD}.
\label{EQN_hE1}
\end{eqnarray}
By high thermal conductivity, the core is believed to be at an essentially uniform temperature. 
As a result, the atmosphere assumes a mean temperature $\xi T_{Edd}$ with $\xi\lesssim \eta$. By (\ref{EQN_hE1}), this implies a height 
\begin{eqnarray}
    H = \xi \,h_E \sim 10\%\, R_{WD}.
    \label{EQN_hE2}
\end{eqnarray}
This expected height (Fig. \ref{fig6}) introduces a radial expansion qualitatively consistent with the data (Fig. \ref{fig:3}).

In this light, the relatively modest characteristic temperature $T_0=1\sim2$\,eV in (\ref{EQN_1b}-\ref{EQN_1c}) can be identified with the core temperature higher by a factor of $T_c/T\simeq \left(T_{Edd}/T\right)^2 =  {\cal O}\left(10^3\right)$ based on Figs. \ref{figEDD}-\ref{fig.1} and \S2. 

The heat transport from core to surface in such atmospheres gives rise to a complex mass-radius relationship, here described by (\ref{EQN_1b}-\ref{EQN_1c}). This appears to be particular relevant to low-mass WDs, essentially below the mean of the WD mass distribution. Above, the relatively high-mass WDs appear to follow the theoretical mass-radius relation (\ref{EQN_1}), evidencing the absence of an atmosphere and/or a relatively low temperature core. 
The origin of this discrepancy appears to be beyond the present considerations, that might involve a distinct composition and associated formation history. Derived for WDs of type DA, our results may serve as a reference for similar model-independent analysis of WDs of different type. 

\section*{Acknowledgements}
The authors thank the anonymous reviewer for constructive comments which greatly contributed to clarity of presentation, and thank M.A. Abchouyeh for stimulating discussions. This work is supported, in part, by NRF No. 
RS-2024-00334550.

\appendix



\bibliographystyle{elsarticle-harv} 
\bibliography{example}






\end{document}